\documentclass[]{elsarticle}
\usepackage[utf8]{inputenc}
\usepackage{float}
\usepackage{amsmath}
\usepackage[hyphens]{url}
\usepackage{hyperref}
\usepackage{booktabs}
\usepackage{caption}
\usepackage{adjustbox}
\usepackage{siunitx}
\hypersetup{breaklinks=true}
\hypersetup{hidelinks=true}

\begin{document}
\author{Carson Ezell\corref{cor1}\fnref{fn1}}
\ead{carson.ezell@cfa.harvard.edu}

\author{Abraham Loeb\fnref{fn1}}
\ead{aloeb@cfa.harvard.edu}

\cortext[cor1]{Corresponding author}
\fntext[fn1]{These authors contributed equally to this work.}
\address{Department of Astronomy, Harvard University, 60 Garden St, Cambridge, Massachusetts 02138, United States}

\bibliographystyle{elsarticle-num-names}

\bibliographystyle{elsarticle-num-names}

\begin{abstract}
The previous decade saw the discovery of the first four known interstellar objects due to advances in astronomical viewing equipment. Future sky surveys with greater sensitivity will allow for more frequent detections of such objects, including increasingly small objects. We consider the capabilities of the Legacy Survey of Space and Time (LSST) of the Vera C. Rubin Observatory to detect interstellar objects of small sizes during its period of operation over the next decade. We use LSST's detection capabilities and simulate populations of interstellar objects in the range of 1-50m in diameter to calculate the expected number of small interstellar objects that will be detected. We use previous detections of interstellar objects to calibrate our object density estimates. We also consider the impact of the population's albedo on detection rates by considering populations with two separate albedo distributions: a constant albedo of 0.06 and an albedo distribution that resembles near earth asteroids. We find that the number of detections increases with the diameter over the range of diameters we consider. We estimate a detection rate of up to a small ISO every two years of LSST's operation with an increase by a factor of ten for future surveys that extend a magnitude deeper. 
\end{abstract}

\begin{keyword}
interstellar objects \sep astronomical instrumentation 
\end{keyword}

\title{Detection Rate of $<$50-meter Interstellar Objects with LSST}

\maketitle

\newpageafter{abstract}

\section{Introduction}

The NASA Authorization Act of 2005 included a mandate—known as the George E. Brown Jr. mandate—to “detect, track, catalog, and characterize the physical characteristics of near-Earth objects equal  or greater than 140 meters in diameter in order to assess the threat of such near-Earth objects to the Earth” \citep{united_states_code_51_2022}. Fulfilling the mandate has served as a guiding principle for astronomical instruments tracking objects in the solar system since it was created. As a result, less attention has historically been paid to objects smaller than $140$ m in diameter as well as rare interstellar objects (ISOs) when designing equipment, data processing pipelines, or conducting data analysis. While surveys of the solar system would have catalogued any such objects that they detected,  most previous sky surveys would have been unable to detect small objects, which we define as objects in the $\leq$50 m diameter range, because of the small region around the Earth in which they would have been observable. 

Over the past decade, the main contributors to cataloging the population of near-Earth objects (NEO) and potentially hazardous asteroids (PHAs) have been the Catalina Sky Survey \citep{larson_current_2007} and Pan-STARRS project \citep{denneau_pan-starrs_2013}. The Legacy Survey of Space and Time (LSST) on the Vera C. Rubin Observatory in Chile will be more sensitive than previous telescopes, resulting in significant scientific implications. Importantly for NASA, LSST will increase the completeness of the catalog of NEOs. Numerous studies have estimated that LSST will identify between 62 percent and 75 percent of this PHA population using its baseline design \citep{ivezic_lsst_2006, grav_modeling_2016, veres_high-fidelity_2017, jones_large_2018}, and the completeness of the PHA catalog would then reach about 86 percent \citep{jones_large_2018}. 

The increased sensitivity of LSST also suggests that it will improve our ability to detect objects of smaller sizes and at greater distances, including small ISOs. In this paper, we place constraints on LSST's detection capabilities for small ISOs and calculate the expected number of such objects that we expect it to detect over its decade of operation.

The Vera C. Rubin Telescope on which the LSST will be conducted has a diameter of  8.4 m (with a 6.67 m diameter mirror), providing a 9.6 deg$^2$ field of view \citep{ivezic_lsst_2006} with a 3.2 billion pixel camera that surveys the entire southern sky every 4 days. Taking an inventory of the solar system is one of the four main science themes of the LSST design \citep{ivezic_lsst_2019}. Previous studies have estimated the rate of detection of interstellar objects by LSST, but these studies have focused on ‘Oumuamua-like objects \citep{seligman_feasibility_2018,hoover_population_2022} or asteroids and comets with a diameter greater than 1 km \citep{engelhardt_observational_2017}. \citet{cook_realistic_2016} considered a variety of objects, including interstellar asteroids as small in diameter as $20$ m. However, these estimates were made prior to the detection of several interstellar objects in as small as the $\sim$ 1 m range, which have led to updated estimates of ISO density \citep{siraj_interstellar_2022}. Previous studies also do not characterize the detection capability of astronomical surveys for objects $\leq$10 m in diameter, which may be observable under particular conditions. We do not consider objects under $5$ m in diameter in our analysis because they must pass very close to Earth to be observed, leading to significant trailing losses that render them undetectable \citep{cook_realistic_2016}.

There would be important scientific benefits to detecting small ISOs, especially given the scarcity of previous ISO observations altogether. Over the past decade, the first four interstellar objects (ISOs) have been discovered. The first interstellar object reported was ‘Oumuamua, which had a diameter of about $200$m and was detected by the PAN-STAARS survey \citep{meech_brief_2017}. The first interstellar meteor discovered (IM1) had a diameter of about $0.45$m \citep{siraj_2019_2022}, and the second interstellar meteor (IM2) displayed similar physical characteristics and was about twice as large as IM1 \citep{siraj_interstellar_2022}. However, both interstellar meteors were detected upon entering the atmosphere of the Earth and burning up, rather than being observed by sky survey equipment. Finally, the interstellar comet 2I/Borisov was detected in 2019 and was imaged by the William Herschel Telescope and Gemini North Telescope \cite{guzik_initial_2020}. Further detections would enable an improved characterization of the ISO population, including the distribution of their sizes, albedos, and densities. Such insights could allow us to place further constraints on our explanations for the origin of ISO populations. The observed interstellar meteors have also demonstrated interesting properties of scientific interest, including unusual material strength \citep{siraj_interstellar_2022}. Furthermore, small ISOs could be evidence of extraterrestrial equipment, which may have gone undetected by previous sky surveys if the objects did not remain near the earth for extended periods of time and had relatively low albedos.

\section{Methodology}

Our calculation proceeds as follows. First, we determine the maximum distance at which an ISO is detectable by LSST as a function of its diameter, albedo, and phase angle. Then, we can determine the size of the observable region for objects with a given diameter and albedo. Next, we calculate the distribution of object chord lengths through the observable region and the distribution of ISO velocities relative to the Earth to estimate the proportion of objects crossing through the observable region that will be detected by LSST. Then, we estimate the crossing rate of ISOs through the observable region based on their inferred density to determine the rate at which we expect to detect such objects. Finally, we show how the estimated detection rate of objects with LSST varies by object diameter and albedo, where the rate of detection for smaller objects is lower than larger objects despite the higher number density of smaller objects. We also find that the detection of small ISOs is sensitively dependent on our assumed albedo distribution, and the number of detections drops significantly if most ISOs have an albedo near 0.06. 

The distance out to which an object is detectable from Earth depends upon its brightness, which we can calculate from its diameter, albedo, and phase angle. The absolute magnitude of objects can be determined as,

\begin{equation}
    H = -5 \log_{10}(\frac{D\sqrt{p_v}}{1329}),
\end{equation}
where $D$ is the diameter (km) and $p_v$ is the albedo \citep{bowell_application_1989}. 

The albedo distribution of ISOs is largely unknown because we lack a significant quantity of previous observations. We consider the case of the ISO population having a constant albedo of 0.06. We also consider the case of the albedo distribution of ISOs being similar to near earth asteroids. To create the latter distribution, for each object in our sample, we draw an albedo from the bimodal Rayleigh distribution described in \citet{wright_albedo_2016}, where,

\begin{equation}
    p(p_v) = f_D\frac{p_v e^{-p_v^2/2d^2}}{d^2}+(1-f_D)\frac{p_ve^{-p_v^2/2b^2}}{b^2}.
\end{equation}

We use the same parameters which were found by \citet{wright_albedo_2016} where $f_D=0.253$, $d=0.03$, and $b=0.168$. The distribution reflects that the Wide-field Infrared Survey Explorer (WISE) found two populations of asteroids classified by albedo where $d$ is the peak albedo of the darker population, $b$ is the peak albedo of the brighter population, and $f_D$ is the proportion of asteroids in the darker population \citep{wright_albedo_2016}. We assume that small interstellar objects are not comets because comets of small sizes entering the solar system would evaporate before they could make a close approach to Earth \citep{levison_mass_2002}. Thus, cometary tails would be more easily detectable at large distances and are ignored here.

We can calculate the absolute magnitude of ISOs if we know their albedo and diameter. To determine the apparent magnitudes of interstellar meteors, we use a similar process to \citet{seligman_feasibility_2018} and \citet{hoover_population_2022}. The apparent magnitude can be calculated from the absolute magnitude as, 

\begin{equation}
    m = H + 2.5\log_{10}(\frac{d_{BS}^2d_{BO}^2}{p(\Theta)d_{OS}^4}),
\end{equation} where $d_{BS}$ is the distance from the body to the Sun, $d_{BO}$ is the distance from the body to the observer—in this case Earth—and $d_{OS}$ is the distance from the observer to the Sun. The phase integral, $p(\Theta)$, used in equation (3) is calculated as,

\begin{equation}
    p(\Theta) = \frac{2}{3}((1 - \frac{\Theta}{\pi})\cos(\Theta) + \frac{1}{\pi}\sin\Theta),
\end{equation} where the phase angle, $\Theta$, can be calculated from the previously defined distances where,

\begin{equation}
    \cos\Theta = \frac{d_{BO}^2+d_{BS}^2-d_{OS}^2}{2d_{BO}d_{BS}}.
\end{equation}

For a given phase angle, $\Theta$, we can then calculate the maximum distance at which an object of a given absolute magnitude would be observable. We can then approximate the total cross-sectional area of the observable region with the ecliptic as,
\begin{equation}
    A = \frac{\pi}{N}\sum\limits_n(d_{BO}^n)^2,
\end{equation}
where $N$ is the number of samples, $n \in [0,\pi]$ are the phase angles and are uniformly spread across the entire interval, and $d_{BO}^n$ is the maximum distance at which an object is observable at phase angle $n$. 

Figure 1 shows the dependence of the maximum distance from the Earth at which an object is detectable on the phase angle for objects of various diameters, given an albedo of 0.06. Figure 2 shows the cross-section of the detectable region of objects with the ecliptic for objects with various diameters and an albedo of 0.06.

\begin{figure}[H]%
\centering
\includegraphics[width=0.75\textwidth]{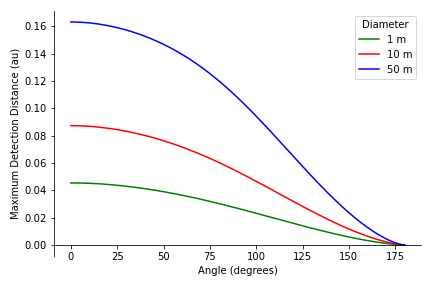}
\caption{The maximum distance at which an ISO with an albedo of 0.06 can be detected as a function of the phase angle between the Sun, object, and Earth for various ISO diameters.}\label{fig1}
\end{figure}

\begin{figure}[H]%
\centering
\includegraphics[width=0.75\textwidth]{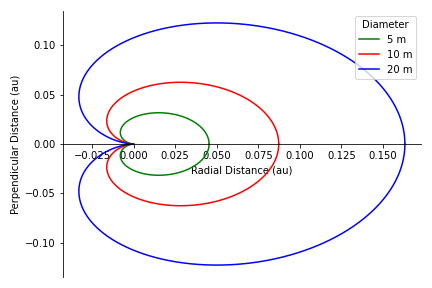}
\caption{Cross-sectional area of the detection region for ISOs with an albedo of 0.06 in the ecliptic plane for various diameters. The plot is centered on Earth where the radial direction (x-axis) points away from the Sun and the perpendicular direction (y-axis) is tangential to the Earth’s position vector relative to the Sun.}\label{fig1}
\end{figure}

Next, we consider the velocity distribution of ISOs with respect to Earth. We assume that ISOs have the same velocity dispersion as local stars around the LSR of $(\sigma_U, \sigma_V, \sigma_W) = (33 \pm 4, 38 \pm 4, 23 \pm 2)$ km s$^{-1}$, and we generate a population of 100,000 ISOs according to this distribution. The velocity distribution of the objects does not depend on the object albedo or diameter. 

The velocity distribution of the objects relative to the Earth depends on the  direction of the Earth’s velocity with respect to the Sun. For each ISO, we draw a corresponding velocity vector for the Earth by selecting a random position within its orbit. The velocity of the Earth for each draw is given in ecliptic coordinates as $(v_x,v_y,v_z) = (v_{\oplus}\cos\theta,v_{\oplus}\sin\theta,0)$ where $\theta \in [0, 2 \pi]$. We also assume the magnitude of the Earth’s velocity is constant at $v_{\oplus}=30$ km s$^{-1}$. 

We draw the velocity vector for the Earth in ecliptic coordinates relative to the Sun for simplicity, but the velocity vectors for the population of interstellar objects are given in galactocentric coordinates relative to the LSR. As a result, we must convert the velocity of the Earth to the galactic coordinate system relative to the LSR, where we use the procedure outlined in \citet{mccabe_earths_2014}. Further details about this procedure are given in Appendix A. 

As the object passes through the solar system, its velocity will increase because of gravitational focusing from the Sun and the Earth. To account for gravitational focusing from the Sun, we multiply the velocity vectors of the interstellar objects by a factor of $\sqrt{v_{\infty}^2+v_{esc}^2}/v_{\infty}$,
where $v_{esc}$ is the escape velocity from the solar system at 1 au from the Sun. The escape velocity is calculated as,

\begin{equation}
    v_{esc} = \sqrt{2 GM_{\odot} / d_{\oplus}},
\end{equation} where $M_{\oplus}$ is mass of Sun and $d_{\oplus}$ = 1 au is the Earth-Sun separation. We repeat the same procedure to calculate the effect of gravitational focusing from the Earth. We find that effects from gravitational focusing from the Earth are negligible because most objects in the population are too far from the Earth for its gravitational focusing effect to be significant, but we nevertheless include this adjustment in our calculation. If a similar procedure to ours is used in the future to calculate the detection rates of very small objects near the Earth by more sensitive telescopes, the effect of gravitational focusing from the Earth may be more significant. In Figure 3, we show the differential probability density function (PDF) of the velocity relative to the Earth of our population of ISOs.

\begin{figure}[H]%
\centering
\includegraphics[width=0.75\textwidth]{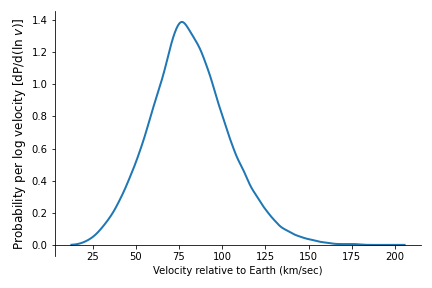}
\caption{The probability per logarithmic bin in velocity, where $v$ is the velocity relative to the Earth for the sample population of ISOs. The minimum velocity relative to the Earth is equivalent to the escape speed from the Earth. We assume the velocity distribution is ISOs is similar to local stars around the LSR and is independent of other variables, such as albedo or diameter.}\label{fig1}
\end{figure}

LSST will survey the sky about 100 times per year, resulting in a complete survey of the sky every 3.65 days \citep{ivezic_lsst_2006}. Objects must be detected by LSST three times to form a track for the object to be recorded, so objects need to remain within a detectable region for about 10.95 days. Since small ISOs must pass close to Earth to be detectable, they may pass through the detectable region in less than 10.95 days despite being within the lower tail for velocity relative to the Earth. Nevertheless, a sufficiently high density of small ISOs might allow for the detection of many such objects that have a relatively low velocity and long chord length through the detectable region.

The amount of time that an object remains visible depends upon its chord length through the observable region. However, since the maximum observable distance depends upon the phase angle, the observable region is not spherical (as shown in Figure 2). Furthermore, the size of the observable region is different for each object since it depends on the object's albedo. We can generate a distribution of chord lengths from our population by randomly assigning each object two points on the boundary of the observable region and calculating the distance between them.

 At each possible phase angle between $0^{\circ}<\Theta<180^{\circ}$, the maximum distance of the observed object is fully constrained for a given diameter and albedo. As a result, given the phase angle, the boundary of the observable region is a circle perpendicular to the ecliptic plane and centered on the vector passing through the center of the Earth and the Sun. The boundary circle is a cross-section of the boundary of the observable region formed by constraining the phase angle. For phase angles $0^{\circ}<\Theta<90^{\circ}$, the object would be in opposition, so the cross-sectional circle would be on the side of the Earth opposite the Sun. For phase angles $90^{\circ}<\Theta<180^{\circ}$, the cross-sectional circle would be between the Earth and the Sun. At each phase angle, the radius of the cross-sectional circle also varies, where it has a radius of $0$ when $\Theta = 0^{\circ}$ or $\Theta=180^{\circ}$. Using the law of sines, we can calculate the radius of the cross-sectional circle given the phase angle as,
\begin{equation}
    r = \frac{d_{bo}d_{bs}\sin(\Theta)}{d_{os}}.
\end{equation} 

When we draw our random sample of boundary points to generate chords through the region and the distribution of chord lengths, we select each phase angle with a probability in proportion to the radius of the cross-sectional circle which demarcates the boundary of the observable region at that phase angle. We calculate a list of $N$ radii which each corresponds to a separate phase angle, $n$, which are uniformly distributed over the interval $0^{\circ} < \Theta <180^{\circ}$. We let $r_n$ be the radius of the circle demarcating the boundary of the observable region at phase angle $n$. Figure 4 shows the relationship between the radius of the observable region for ISOs and the phase angle. Then, the probability that an interstellar object enters or leaves the observable region at a phase angle closest to $\Theta_n$ for any $n$ is,
\begin{equation}
    P(\Theta_n) = \frac{r_n}{\sum\limits_n r_n}.
\end{equation}

We draw phase angles from the resulting probability distribution and assign one to each boundary point. Given the phase angle, we can calculate the value $d_{bo}$, which is constrained since the object is at the maximum observable distance for the phase angle. Each boundary point we draw must also be randomly located on the cross-sectional circle corresponding to its phase angle. Thus, for each boundary point, we also randomly draw another angle, $\phi$, from the uniform distribution $\sim U[0,2\pi]$, which corresponds to the angle of the object within the cross-sectional circle. Then, we then let the position of the boundary point in ecliptic coordinates be $(d_{bo}^2-r_n^2,r\cos\phi,r\sin\phi)$. Here, the x-direction is the Sun-Earth vector along the ecliptic, the y-direction is the perpendicular vector within the ecliptic plane, and the z-direction is perpendicular to the ecliptic plane.

\begin{figure}[H]%
\centering
\includegraphics[width=0.75\textwidth]{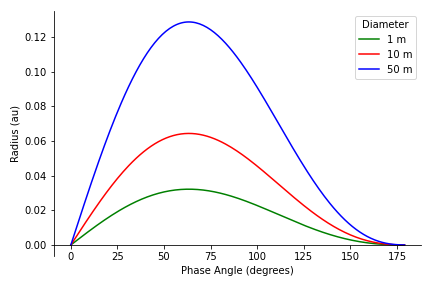}
\caption{Radius of the cross-sectional circle which demarcates the boundary of the region in which ISOs can be detected. We plot the phase angle on the x-axis, assume an albedo of 0.06, and plot the curve for three separate diameters.}\label{fig1}
\end{figure}

\begin{figure}[H]%
\centering
\includegraphics[width=1\textwidth]{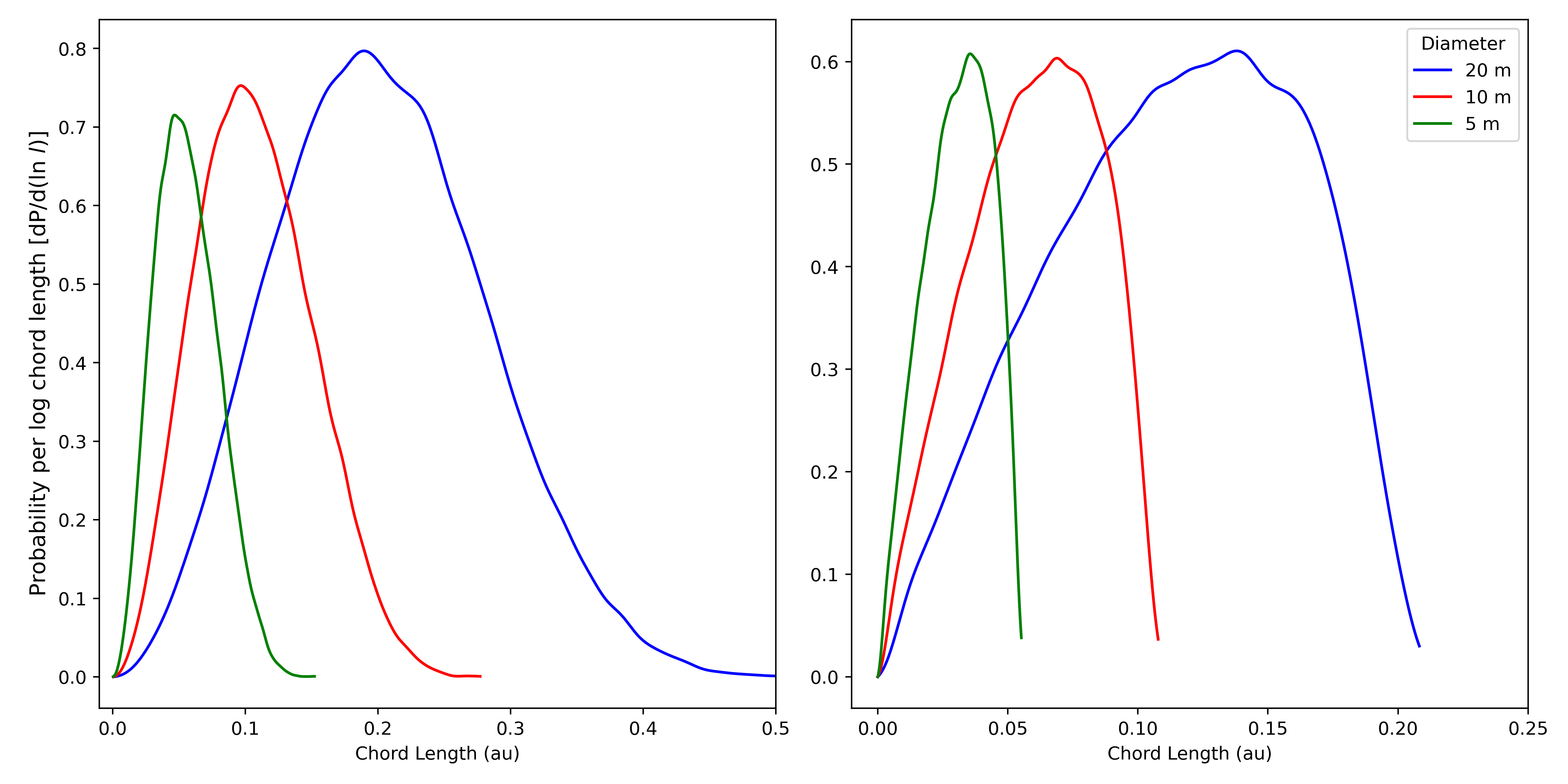}
\caption{The probability per logarithmic bin in chord length, where $l$ is the chord length through the observable region, for the distribution of chord lengths for ISO populations with various diameters. Two albedo distributions are depicted: the albedo distribution from equation (2) on the left and a population with a constant albedo of 0.06 on the right.}\label{fig1}
\end{figure}

We choose pairs of two boundary points and determine the chord length through the observable region between them. We pair each chord with an ISO in a sample population and determine its distance covered relative to the Earth over a period of 11 days. If the total distance covered is less than the chord length, we consider the object to meet the criteria for detectability. Figure 5 shows the differential PDF of chord lengths for ISO populations with various diameters and albedo distributions. The albedo affects the distance of points on the boundary region from the Earth, where objects with higher albedos are visible at greater distances. 

Next, we use past detection rates of ISOs to calibrate our estimate for the number density of ISOs. ISOs on the order of $\leq 1$m have been detected as the IM1 and IM2 fireballs burning up in the atmosphere of the Earth. If we assume that we detect all ISOs which enter the atmosphere of the Earth, we can estimate the passing rate of ISOs of similar diameters to those objects. Then, we can calculate the number density of such objects as,

\begin{equation}
    n \simeq \frac{\Gamma}{v_{\infty}\pi R_{\oplus}^2[1+(v_{esc}/v_{\infty})^2]},
\end{equation} which was demonstrated in \citet{siraj_interstellar_2022}. The implied rates of detection for IM1 and IM2 are  $\Gamma \sim 0.1$ yr$^{-1}$ \citep{siraj_2019_2022,siraj_interstellar_2022}. The resulting number density for IM1-like and IM2-like objects is about $n = 2 \times 10^6$ au$^{-3}$. The corresponding local mass density for IM1-like objects is 1.2 $M_{\oplus}$ pc$^{-3}$, which is an order of magnitude lower than the estimated local mass density for IM2. We use the given value as the estimate for mass density of IM1-like objects, and we assume a power law size distribution of ISOs with an equal total mass of objects per object log mass bin. This assumption is an approximate estimate that holds across a large distribution of ISO masses \citep{siraj_interstellar_2021}. IM1 has a diameter of about 0.45m, which we use to constrain the relationship \citep{siraj_2019_2022}. We assume that ISOs have a similar density independent of size, so we consider the resulting relation $m \sim D^3$. Let $\Gamma_{IM1}$ be our empirical detection rate of IM1-like interstellar meteors determined to which we calibrate our estimate. Then, our rate of detection of interstellar meteors of a given diameter $D$ with LSST can be calculated as,

\begin{equation}
    \Gamma_D = \Gamma_{IM1}(\frac{D}{45 \text{cm}})^{-3}(\frac{A_D}{A_{\oplus}})p_D,
\end{equation} where $A_D$ is the cross-sectional area of the observable region of interstellar objects of diameter $D$ with LSST, $A_{\oplus}$ is the cross-sectional area of the Earth, and $p_D$ is the detection rate of interstellar objects with diameter $D$. We can use this relationship to estimate the LSST detection rate of interstellar objects with various diameters. However, the estimate becomes less accurate for larger objects, such as those which are ‘Oumuamua-sized ($200$m), because we are extrapolating farther from our observed number density of smaller objects. For these objects, the detection rate of an object larger than IM1, such as ‘Oumuamua, can be used instead of $\Gamma_{IM1}$. However, we do not concern ourselves with 'Oumuamua-sized objects for the purpose of this analysis of small ISOs.

There are also further limitations of the viewing capabilities and survey strategies of LSST that will lead to the detection of less interstellar objects. For example, LSST will only detect interstellar objects within its field-of-view (FoV), which is determined by its survey strategy. The Wide Fast Deep survey (WFD), which will take up about 90 percent of the observing time for LSST, will include declinations from $-60^{\circ}$ to $5^{\circ}$, including all right ascension values \citep{jones_large_2018}. Hence, LSST will primarily detect objects that fall within this range, which is about $18,000 \text{ deg}^2$ \citep{bianco_optimization_2021}. We update our estimated number of detections by the proportion of the sky within the FoV of the WFD survey.

Above, we outlined a procedure to calculate the estimated number of ISO detections with LSST given an object diameter, velocity distribution, and albedo distribution. We repeat this calculation using various diameters and our two albedo distributions for ISOs. We consider objects between 1m and 50m in diameter, and we find that the estimated number of objects detected grows exponentially with diameter in this range. This is consistent with our expectation that smaller objects will not be detectable by LSST because they will not remain within the observable region for a sufficient amount of time. For an albedo distribution based on equation (2), we estimate the detection of about $5-6$ ISOs with a diameter $\leq 50$m in the decade of LSST's operation. Given a constant albedo of 0.06 throughout the population, we estimate the detection of $1-2$ ISOs with a diameter $\leq 50$m in the same period of operation.

Our estimates for the number of ISOs detected by LSST of various diameters, given an albedo distribution, are shown in Table 1. Our estimates are relatively aligned with those of \citet{hoover_population_2022}, which estimated about 15 'Oumuamua-sized objects ($\sim 200$-m diameter) in LSST's decade of operation. Our estimate is about twice the estimate of \citet{cook_realistic_2016} for a constant albedo population, which may reflect increased ISO density estimates following the detection of the first four ISOs.

\begin{table}[]
    \centering
    \captionsetup{width=0.9\textwidth} 
    \begin{adjustbox}{width=\textwidth, center}
    \sisetup{table-format=2.2, table-number-alignment=center}
    \begin{tabular}{S[table-format=2.0] S S}
        \toprule
        {Diameter (m)} & {Detection Rate (\citet{wright_albedo_2016} albedo distribution)} & {Detection Rate (albedo = 0.06)} \\
        \midrule
        1  & 0    & 0 \\
        5  & 0.1  & 0 \\
        10 & 0.42 & 0 \\
        20 & 3.40 & 0.57 \\
        30 & 7.63 & 2.03 \\
        40 & 12.45 & 3.94 \\
        50 & 15.53 & 6.17 \\
        \bottomrule
    \end{tabular}
    \end{adjustbox}
    \caption{The estimated detection rate (objects per decade) for small interstellar objects by LSST based on our simulated results.}
    \label{tab:my_label}
\end{table}

Some of our estimates for smaller ISOs might be overestimates because of the trailing loss—the objects spend most of their time in the observable region in very close proximity to the Earth. This effect is less substantial for larger objects because their angular velocity through the sky is small in most of the observable region, which is sufficiently far from the Earth. However, such objects may become undetectable because of trailing loss when very close to the Earth—if one of the three observations occurs at such a time, a track of the object may not be formed.

\begin{figure}[H]%
\centering
\includegraphics[width=1\textwidth]{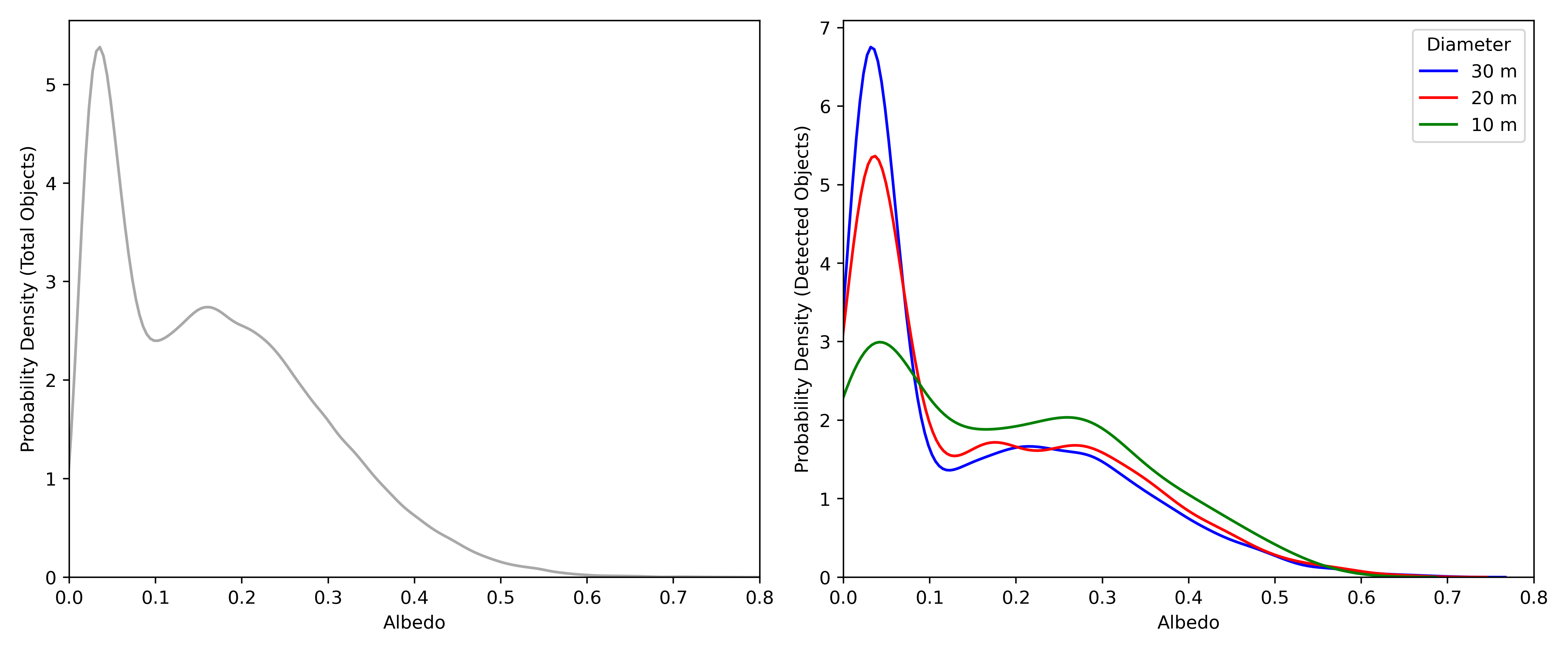}
\caption{Probability density functions of albedo distributions of entire sample interstellar object population (left) and detected interstellar object populations for various diameters (right). The populations are constructed from albedo distributions which resemble that of near-earth asteroids.}\label{fig1}
\end{figure}

In Figure 6, we show the differential probability for the albedos of detected ISOs within the sample population that follows the albedo distribution from equation (2), and we also show the differential probability of the original sample population. We find that the high albedo population allows for significantly more detections, especially for smaller ISOs. If the true albedo distribution of ISOs were similar to that of near earth asteroids as suggested by this sample population, we would likely have seen more ISO detections with previous telescopes, suggesting this is unlikely to be the albedo distribution of interstellar objects. However, interstellar objects are likely to originate from different populations, and we cannot rule out that some populations may contain objects with higher albedos like the near earth asteroids population. Further detections of ISOs can allow us to produce better estimates of their albedo distribution, including distinguishing between populations.

We also consider the extent to which the detection rate is impacted by changes to the sensitivity of the telescope. In Figure 7, we show the impact of improvements in telescope sensitivity on the detection rate for objects of various diameters. A telescope that can detect all objects where the limiting magnitude $m \leq 25$ can detect approximately an order of magnitude more objects in the $50$ m-diameter range than LSST, which can only detect objects where $m \leq 24$.

\begin{figure}[H]%
\centering
\includegraphics[width=0.99\textwidth]{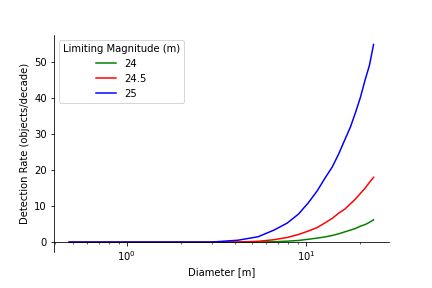}
\caption{Detection rates of ISOs of various diameters given the limiting magnitude (higher numbers are fainter) at which an object can be detected. LSST has a minimum magnitude of $m=24$ (shown in green). Our plot shows estimates for the sample population where all objects have an albedo of 0.06. }\label{fig1}
\end{figure}

\section{Conclusion}
Our calculations forecast that LSST will be able to detect a small ISO in the diameter range of 1-50m every 1-2 years. Extending the sensitivity of the survey by one magnitude will increase this number by a factor of ten. 

While solar system surveys have historically been of scientific and political interest because of the desire to catalog the NEO and PHA populations—including for planetary protection—the emerging ability to detect ISOs and smaller objects will lead to new scientific research opportunities. Furthermore, we expect future telescopes to continue the trend of enabling the detection of even smaller objects. 

In addition to ISOs, very sensitive telescopes may be able to detect other objects of economic or commercial interest, including tracking probes in orbit around celestial bodies and in transit through the solar system. Telescopes may also be more equipped to detect rovers and probes on commercial or scientific missions throughout the solar system, small asteroids of interest for a future asteroid mining industry, and weather events or future human space settlement activities on celestial bodies. In the very long-term, space-based astronomical instrumentation may transform geopolitics and scientific research through the ability to make observations from extremely small probes to detailed accounts of rare events \citep{scherrer_ultra_2023}. When assessing the ability of our instruments to detect small objects besides ISOs, the same assumptions cannot be made about the albedos and position and velocity distributions of such objects, but a similar methodology to our own can be used to determine whether the viewing capabilities of an astronomical survey are sufficient for detection.

\appendix
\section{Velocity Conversion}

To compare the velocity vectors of the ISOs to the velocity vector of the Earth, we need to convert the velocity vector of the Earth into galactic coordinates. In addition, the velocity vector of the Earth is given relative to the Sun, and the velocity vector of the ISOs are given relative to the LSR. Thus, we also subtract the peculiar velocity of the Sun from the velocity vector of the ISOs so that we can compare the object’s velocity to that of Earth. We use the procedure outlined in \citet{mccabe_earths_2014}.

First, we convert the velocity vector from ecliptic to equatorial coordinates using the rotation matrix,

$$R = \begin{bmatrix} 1 & 0 & 0 \\ 0 & \cos \epsilon & - \sin \epsilon \\ 0 & \sin \epsilon & \cos \epsilon
\end{bmatrix}$$
    
where $\vec{v}_{\oplus,eq}=R\vec{v}_{\oplus,ec}$ and is the obliquity of the ecliptic plane. Then, we convert the velocity vector from equatorial to galactic coordinates through the relation $\vec{v}_{\oplus,gal}=M\vec{v}_{\oplus,eq}$ where,
$$M_{11}= -\sin l_{CP} \sin \alpha_{GP}- \cos l_{CP} \cos \alpha_{GP} \sin \delta_{GP}$$
$$M_{12}= \sin l_{CP} \cos \alpha_{GP}- \cos l_{CP} \sin\alpha_{GP} \sin \delta_{GP}$$
$$M_{13}= \cos l_{CP} \cos \delta_{GP}$$
$$M_{21}= \cos l_{CP} \sin\alpha_{GP}-\sin l_{CP} \cos\alpha_{GP} \sin\delta_{GP}$$
$$M_{22}= -\cos l_{CP} \cos\alpha_{GP}-\sin l_{CP} \sin\alpha_{GP} \sin\delta_{GP}$$
$$M_{23}= \sin l_{CP} \cos\delta_{GP}$$
$$M_{31}= \cos\alpha_{GP} \cos\delta_{GP}$$
$$M_{32}= \sin\alpha_{GP}\cos\delta_{GP}$$
$$M_{33}= \sin\delta_{GP}$$

Since we have the velocity of the Earth relative to the Sun, we then subtract the peculiar vector of the Sun relative to the LSR $\vec{v}_{pec}=(v_U,v_V,v_W) = (10 \pm 1, 11 \pm 2, 7 \pm 0.5)$ km s$^{-1}$ from the velocity vectors of the interstellar meteors to find the objects’ velocity vectors relative to the Sun in the galactic coordinate system,
\begin{equation} \vec{v}_{o,Sun}=\vec{v}_{o,LSR}-\vec{v}_{pec},
\end{equation} 
where $v_{o,Sun}$ is the object’s velocity relative to the Sun, $v_{o,LSR}$ is the object’s velocity relative to the LSR, and $v_{pec}$ is the velocity of the Sun relative to the LSR  from the velocity vectors of the ISO population drawn from the LSR. Then, we can calculate the velocities of the ISOs relative to the Earth as,

\begin{equation}
v_e = || \vec{v}_{o,Sun} - \vec{v}_{\oplus, gal}||.
\end{equation}

\bibliography{lsstdetectionrate}
\end{document}